\documentclass{mn2e}
\usepackage{epsfig}
\def\msun{{\rm M_{\odot}}}

\def\ngc{{NGC 4051}}

\def\msun{{\rm M_{\odot}}}

\def\xmm{{\it XMM-Newton}}
\def\chandra{{\it Chandra}}

\def\et{{et al.\ }}

\def\suzaku{{\it Suzaku}}

\newcommand{\ls}{\mathrel{\hbox{\rlap{\hbox{\lower4pt\hbox{$\sim$}}}\hbox{$<$}}}}
\newcommand{\gs}{\mathrel{\hbox{\rlap{\hbox{\lower4pt\hbox{$\sim$}}}\hbox{$>$}}}}

\def\H0{{\rm ~km~s^{-1}~Mpc^{-1}}}

\def\msun{M_{\rm \odot}}

\def\et{{et al.}}

\title[Shocked outflow in \ngc]
{The shocked outflow in \ngc\ -- momentum-driven feedback, UFOs and warm absorbers}
\author[K.A.Pounds \et]
        {K.A.Pounds and A.R.King 
	\\
Department of Physics and Astronomy, University of Leicester,
Leicester, LE1 7RH, UK\\}

\date{Accepted ; Submitted }
\pagerange{\pageref{firstpage}--\pageref{lastpage}}
\pubyear{2010}
\begin{document}
\maketitle
\label{firstpage}

\begin{abstract}  
 
An extended \xmm\ observation of the Seyfert 1 galaxy \ngc\ in 2009 revealed an unusually rich absorption spectrum with outflow velocities, in both
RGS and EPIC spectra, up to $\sim$ 9000 km s$^{-1}$ (Pounds and Vaughan 2011). Evidence was again seen for a fast ionised wind with velocity
$\sim$0.12c (Tombesi 2010, Pounds and Vaughan 2012). Detailed modelling with the XSTAR photoionisation code now confirms the general correlation of
velocity and ionisation predicted by mass conservation in a Compton cooled shocked wind (King 2010). We attribute the strong column density gradient
in the model to the addition of strong two-body  cooling in the later stages of the flow, causing the ionisation (and velocity) to fall more
quickly, and confining the lower ionisation gas to a narrower region. The column density and recombination timescale of the highly ionised flow
component, seen mainly in Fe K lines, determine the primary shell thickness which, when compared with the theoretical  Compton cooling length,
determines a shock radius of $\sim$$10^{17}$ cm.  Variable radiative recombination continua (RRC) provide a key to scaling the lower ionisation gas,
with the RRC flux then allowing a consistency check on the overall flow geometry. We conclude that the 2009 observation of \ngc\ gives strong
support to the  idea that a fast, highly ionised  wind, launched  from the vicinity of the supermassive black hole, will lose much of its mechanical
energy after shocking against the ISM at a sufficiently small radius for strong Compton cooling.  However, the total flow momentum will be
conserved, retaining the potential for a powerful AGN wind to support momentum-driven feedback (King 2003; 2005). We speculate that the `warm
absorber' components often seen in AGN spectra result from the accumulation of shocked wind and ejected ISM.

\end{abstract}


\section{Introduction}

High resolution spectra of the bright Seyfert 1 galaxy \ngc\, obtained by \chandra, \xmm\ and \suzaku\ over the past decade have detected soft
X-ray  absorption  lines indicating a ubiquitous outflow with velocities in the range $\sim$200-600 km s$^{-1}$  (Collinge \et\ 2001, Ogle \et\
2004,  Pounds \et\ 2004, Steenbrugge \et\ 2009), with occasional reports of higher velocities of $\sim$2340 km s$^{-1}$ (Collinge \et\ 2001) and
$\sim$4600 km s$^{-1}$ (Steenbrugge \et\ 2009). A substantially longer \xmm\ observation in 2009 revealed a more complex soft X-ray absorption
spectrum, with outflow velocities  up to $\sim$9000 km s$^{-1}$ (Pounds and Vaughan 2011). While rare in the soft X-ray band, similar outflow
velocities are frequently observed in Fe K spectra of \ngc, more sensitive to highly ionised matter (Pounds \et\ 2004, Lobban \et\ 2012). Moreover,
in an archival \xmm\ search of 42 radio quiet AGN, Tombesi \et\ (2010) found evidence for still higher velocity outflows (v$\sim$0.1c) in  $\sim$35
$\%$ of their sample, including \ngc. An intriguing question is whether, or how, such ultra-fast outflows (UFOs) are physically linked with the more
weakly ionised and much slower outflows (`warm absorbers') which are also common in many bright AGN.

In an initial analysis of the 2009 \xmm\ observation of \ngc, Pounds and Vaughan  (2011; hereafter Paper I) considered an apparent correlation of
outflow velocity and ionisation parameter in terms of a mass-conserved decelerating flow, perhaps resulting from strong Compton cooling after
shocking of the high speed primary wind with the ISM or slower moving ejecta (King 2010;  Zubovas and King 2012). A wider relevance of the  shocked
flow scenario lies in the transformation of an energetic AGN wind into an outflow where the potential mechanism for galaxy feedback is the momentum
thrust (King 2003, 2005).

The high sensitivity of the 2009 \xmm\ observation also found several strong, blue-shifted and broad emission lines (BEL), all showing evidence of
self-absorption near the line cores (Pounds and Vaughan 2011a; hereafter Paper II). Broad soft X-ray emission lines have  previously been reported
for \ngc (Ogle \et\ 2004, Steenbrugge \et\ 2009), and for several other Seyfert 1 galaxies (Kaastra \et\ 2002, Costantini \et\ 2007, Smith \et\
2007), being interpreted by those authors as an extension of the optical/UV `broad line region', envisaged as optically thick `clouds' circling the
central black hole. Paper II outlined an alternative origin of the broad soft X-ray emission lines in \ngc, arising in a limb-brightened shell of
shocked gas, and noted that self-absorption in the near-orthogonal flow could explain the low velocity absorption component seen across a wide range
of ionisation states.  

In the present paper, we describe in Section 3 the detailed modelling of both RGS and EPIC absorption spectra using the XSTAR photoionisation 
code,
with 5 RGS components confirming a linear correlation of outflow velocity and ionisation parameter. While low velocity/low ionisation  absorption is
only observed in the RGS data, the RGS and EPIC model parameters are in good agreement for the higher velocity/higher ionisation absorption. A 6th
RGS component lies off the main trend, and represents the low outflow velocity observed over a wide range of ionisation.  Intriguingly, a 7th RGS
component has a small but significant nett redshift, which we speculate might be due to shocked gas falling back towards the hole.

Section 4 uses the inter-orbit variability of Fe K absorption, reported in more detail in Paper III, to estimate the recombination time of the more 
highly  ionised flow component (from changes in the FeXXV to FeXXVI ratio) and hence constrain the related particle density, which - with the
corresponding  absorption column density from XSTAR modelling - gives the thickness of the highly ionised post-shock shell. In Section 6 this value
is compared with the theoretical Compton cooling length and used to derive an estimate of the shock radius. 

At some point along the post-shock flow, two-body processes will become important, resulting in additional cooling (Appendix), with the flow temperature
(ionisation and velocity) then falling more quickly.  Section 5 examines strong radiative recombination continua (RRC) emission, with inter-orbit
variability providing a recombination timescale and corresonding mean particle density to allow scaling of this lower ionisation flow component.   

In Section 6 we outline a self-consistent physical picture of a highly ionised, high speed wind, which shocks with the
ISM at a sufficiently small radius for strong Compton cooling in the AGN radiation field to result in most of the mechanical energy in the flow being
lost. The subsequent radial structure of the decelerating post-shock flow is determined by the competing cooling processes, which provide a physical
basis on which to understand the complex X-ray absorption and  emission spectra in the 2009 \xmm\ observation of \ngc.

\section{Observations}

\ngc\ was observed by \xmm\ on 15 orbits between 2009 May 3 and June 15, yielding a total on-target exposure of $\sim$650 ks. The present paper
uses  data from the Reflection Grating Spectrometers, RGS 1 and RGS 2 (den Herder \et\ 2001) and the EPIC pn camera (Str\"{u}der \et\ 2001). 
Detailed X-ray light curves are given in Alston \et\ (2013) while mean RGS flux levels for each satellite orbit are shown in figure 1. Table 1
identifies the spacecraft revolution and respective orbit number, with corresponding observation start and end times. The overall observation was
characterised by a generally high soft X-ray continuum over the first half (apart from orbit 4) with four successive high flux orbits (5-8),
followed by lower mean continuum levels from orbit 9 onward. 

\begin{table}
\caption{\textit{XMM-Newton} 2009 observation summary. The columns list (1) orbit number, (2) spacecraft 
revolution number, (3) observation start date and time, (4) observation end date and time}
\begin{tabular}{c c c c}
\hline
orbit no. & XMM & Observation & Observation  \\
& rev. no. & start & end \\
\hline
1 & 1721 & 05-03 10:22 & 05-03 23:04 \\
2 & 1722 & 05-05 10:16 & 05-05 22:57 \\
3 & 1724 & 05-09 10:02 & 05-09 22:41 \\
4 & 1725 & 05-11 09:55 & 05-11 22:32 \\
5 & 1727 & 05-15 12:36 & 05-15 21:40 \\
6 & 1728 & 05-17 09:35 & 05-17 21:22 \\
7 & 1729 & 05-19 09:29 & 05-19 21:06 \\
8 & 1730 & 05-21 09:22 & 05-21 21:01  \\
9 & 1733 & 05-27 09:02 & 05-27 21:31 \\
10 & 1734 & 05-29 09:09 & 05-29 21:18 \\
11 & 1736 & 06-02 08:43 & 06-02 21:12 \\
12 & 1737 & 06-04 09:55 & 06-04 20:58 \\
13 & 1739 & 06-08 08:34 & 06-08 20:40 \\
14 & 1740 & 06-10 08:16 & 06-10 20:37 \\
15 & 1743 & 06-16 08:23 & 06-16 20:15 \\
\hline
\end{tabular}
\label{tab_obs}
\end{table}

\begin{figure}                                                                                                
\centering                                                              
\includegraphics[width=6.3cm, angle=270]{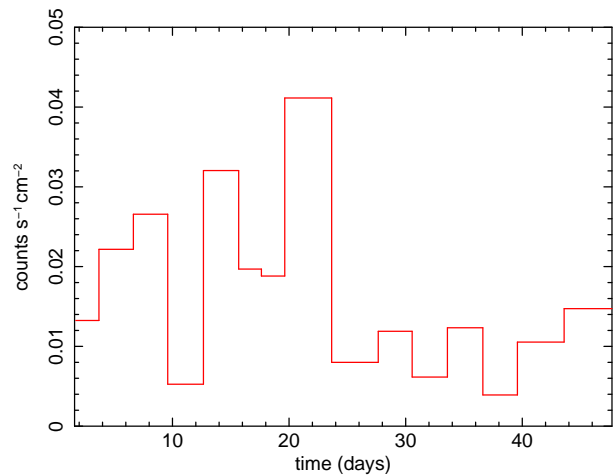}                                                                                  
                                                                               
\caption                                                                
{Mean RGS flux levels for the 15 \xmm\ orbits of the 2009 observing programme. The time axis is in MJD-54952}        
\end{figure}

\section{Modelling the outflow with XSTAR}
To quantify the overall photoionised absorption in the complex outflow in \ngc\ the RGS and EPIC spectra for the sum of the 4 high flux orbits 5-8
were modelled in Xspec (Arnaud 1996) with alternative grids based on the XSTAR photoionisation code (Kallman \et\ 1996). 

\subsection{RGS data} 
For the RGS data, where most absorption lines appear intrinsically narrow, we chose grid 18 from the XSTAR library, which includes a fixed turbulent
velocity of 100  km s$^{-1}$ to minimise computing time. Grid 18 covers an ionisation parameter range in log$\xi$ from -4 to +4 erg cm s$^{-1}$, and
column densities from $10^{19}$ to $10^{23}$ cm$^{-2}$. Abundances of relevant metals from C to Fe were initially set to solar values, but allowed
to vary in the final spectral fitting.  

The composite high-flux RGS spectrum was first fitted by a power law plus black body to provide a smooth match to the continuum. Positive Gaussians
(determined from fitting the sum of the low flux spectra of orbits 4,11,13) were then added to the continuum to represent  the main BEL and the
strong FeXVII emission lines near 17 \AA. A number of {\it redge} components were added to represent the stronger RRC, with the  normalisation of
each left free to vary. Finally, guided by the intriguing observation that very few absorption lines in the RGS spectra appear to penetrate below  $\sim$50$\%$
of the continuum, an unabsorbed fraction of the power law plus black body continuum was included in the model. 

A sequence of photoionised absorbers from grid 18 was then added to achieve the best statistical fit over the whole 6-36 \AA\ RGS waveband.  The
ionisation parameter, column density and velocity (output as a modified redshift) were the primary free parameters of each absorber. In the event
the recorded modelling was limited to the 10-36 \AA\ waveband due to the low RGS sensitivity at shorter wavelengths. 

\begin{figure}
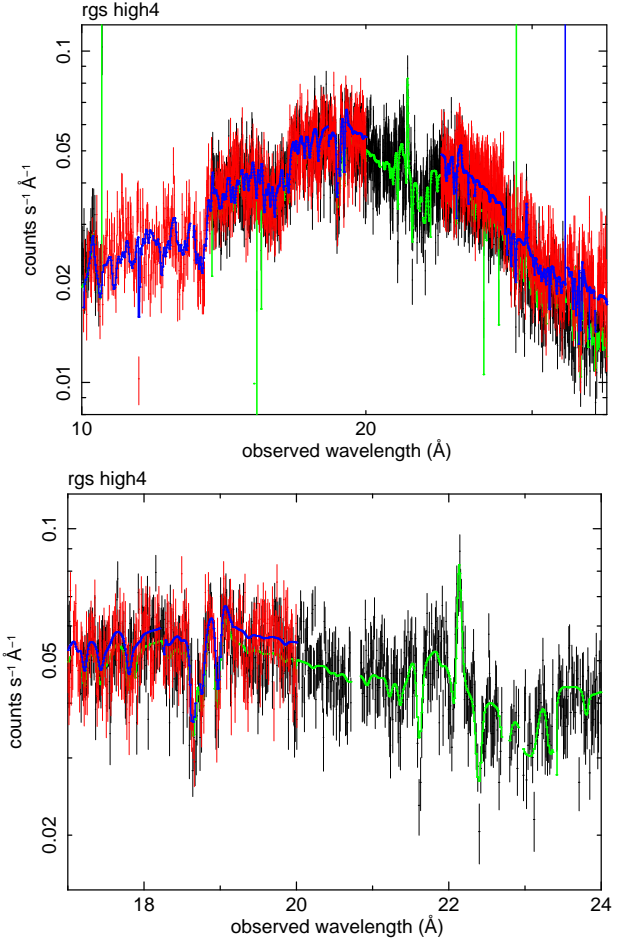
                                                                                                
\centering                                                              
\includegraphics[width=6.2cm, angle=270]{7abs.ps}                                                                                  
\centering                                                              
\includegraphics[width=6.3cm, angle=270]{7abs_a.ps}                                                                                  
\caption                                                                
{XSTAR model fit to the combined RGS1 and RGS2 spectra summed over the four high-flux orbits 5-8 is shown for the full 10-36 \AA\  waveband  in the
upper panel, with the section covering strong absorption in oxygen highlighted in the lower panel}        
\end{figure}

An initial fit to the RGS 1 data over the waveband 17-24 \AA, dominated by absorption lines of OIV, V, VI, VII and VIII and largely independent of
relative abundances, required two photoionised components, expressing the strong low ($\sim$500 km s$^{-1}$) and higher ($\sim$4000 km s$^{-1}$) velocity
absorption in OVII and OVIII. The fit was then extended to 36 \AA, to include the resonance lines of  NVII, NVI and  CVI, and with both RGS1 and
RGS2 data, and finally over the waveband 10-36 \AA, covering the higher energy K-shell resonance transitions of Ne, and a potential complex of Fe-L
lines. For the full band fits the abundances were allowed to vary, although being tied for the same element across the separate ionised components.

A total of 7 photoionised absorbers yielded significant incremental improvements to the 10-36 \AA\ fit, with an overall reduction from
$\chi^{2}$/d.o.f. = 6785/4125 to  $\chi^{2}$/d.o.f. = 5048/4094. 
Intriguingly, a further significant reduction to $\chi^{2}$/d.o.f. = 4996/4093 was
found with 28$\pm$5 \% of the continuum unabsorbed. The parameters of this multi-absorber fit (illustrated in figure 2) are listed in Table 2. 

The abundances of the more important elements in the best fit were C:0.29$\pm$0.16,  N:0.62$\pm$0.35,
O:0.68$\pm$0.34, Ne: 0.41$\pm$0.30 and Fe:1.9$\pm$1.1. While these values are noted, only the over-abundance of Fe made a substantial difference to
the fit parameters.

\begin{figure}
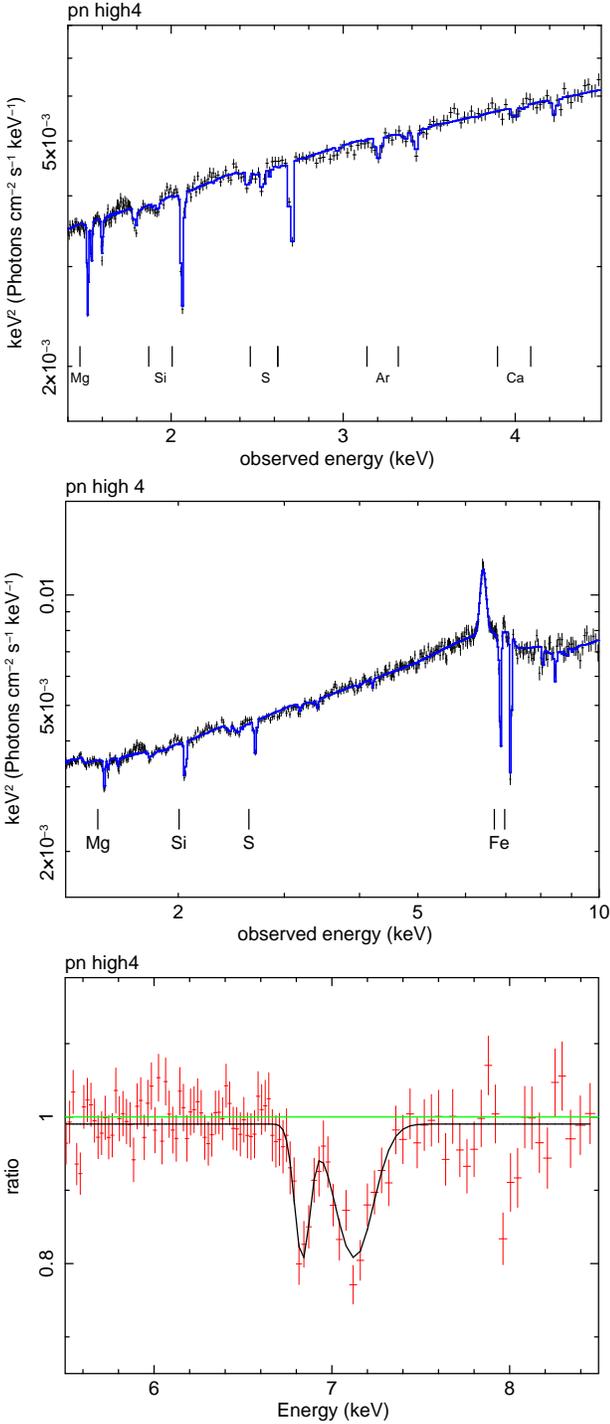
                                                                                               
\centering                                                              
\includegraphics[width=6.3cm, angle=270]{mykonos3a.ps}                                                                                  
\centering                                                              
\includegraphics[width=6.3cm, angle=270]{mykonos4a.ps}                                                                                  
\centering                                                              
\includegraphics[width=6.3cm, angle=270]{mykonos5a.ps}                                                                                  
\caption                                                                
{XSTAR model fits to the combined pn spectra over the four high flux orbits 5-8. The top panel is restricted to the 1.3-5 keV energy
band covering the K-shell resonance lines of Mg, Si, S, Ar and Ca. The mid panel shows a second fit extended to include the strong Fe K
absorption lines. The lower panel reproduces a simple Gaussian fit to the Fe K spectral lines, with a broad  Lyman-$\alpha$ line
indicative of a spread of outflow velocities}        
\end{figure}

Components 1, 2 and 7 represent the high velocity absorption observed most strongly in the soft X-ray data in OVIII and NeX, and it is notable that
they pick out similar  velocities to the values obtained from Gaussian fits to the absorption in OVIII, NVII and CVI Lyman-$\alpha$ (Paper I).
Component 7 is the least significant, with a lower column density indicating a minor constituent at the highest post-shock velocities. 

Components 3 and 4 represent the lower velocity/lower ionisation absorption, observed particulary strongly in absorption lines of OV, OVI and OVII
and the Fe UTA at $\sim$15-16.5 \AA. The absence of an intermediate velocity component
is consistent with the absorption velocity profiles
in Paper I, showing with a relative lack of absorption between $\sim$3000 and $\sim$1000 km s$^{-1}$.

Component 5 represents the low velocity absorption seen across a wide range of ionisation parameter, interpreted in Paper II as self-absorption in
the BEL, where a high-inclination to the line of sight constrains the projected flow velocities. 

Component 6 was a surprise, but is strongly required by the fit. Although only hinted in individual spectra, the requirement of a red-shifted
absorption component at low velocity is interesting as a possible indicator of post-shock matter falling back from the contact  discontinuity,
having slowed to below the local escape velocity. The absence of a similar red-shifted absorber in the comparable analyses of Steenbrugge \et\ (2009)
and Lobban \et\ (2011) would argue against an alternative origin in a high velocity cloud in the line of sight in the host galaxy, as reported for Mkn 509 
(Ebrero \et\ 2011).

\begin{table}
\centering
\caption{Parameters of the photoionised outflow fitted to the RGS data. Column density is in H atoms cm$^{-2}$ and ionisation parameter in erg cm
s$^{-1}$}
\begin{tabular}{@{}lccccc@{}}
\hline
Comp & log$\xi$ & N$_{H}$  & velocity (km s$^{-1}$ & $\Delta$$\chi^{2}$\\
\hline
1 & 3.0$\pm$0.1 & 4.3$\pm$0.4$\times10^{22}$ & -5760$\pm$500 & 488/3 \\
2 & 2.5$\pm$0.1 & 1.0$\pm$0.1$\times10^{22}$ & -3720$\pm$300 & 113/3 \\
3 & 1.6$\pm$0.2 & 7$\pm$0.3$\times10^{20}$ & -510$\pm$150 & 94/3 \\
4 & 0.8$\pm$0.1 & 1.6$\pm$0.3$\times10^{21}$ & -120$\pm$45 & 797/3 \\
5 & 2.6$\pm$0.2 & 2$\pm$0.5$\times10^{21}$ & -630$\pm$350 & 174/3 \\
6 & -0.82$\pm$0.1 & 4$\pm$1.5$\times10^{20}$ & +360$\pm$60 & 48/3 \\
7 & 3.2$\pm$0.2 & 1$\pm$0.3$\times10^{21}$ & -10290$\pm$1000 & 23/3 \\
\hline
\end{tabular}
\end{table}

The outflow velocity and ionisation parameter of components 1-5 and 7 are plotted in figure 4, together with the main absorbing ions located at the
ionisation parameter of maximum abundance for a gas in equilibrium.

\subsection{EPIC data}

Fitting the EPIC pn absorption spectra, again for the sum of the 4 high flux orbits (5-8), allows an extension of the XSTAR modelling to heavier
ions whose K-shell wavelengths fall outside the sensitive range of the RGS.  The generally higher velocities and ionisation levels seen in the EPIC
spectrum led to a preference of grid 21 for XSTAR modelling, with  a higher turbulent velocity of 1000 km s$^{-1}$, but otherwise a similar
parameter range to grid 18. The continuum was first modelled by a partially covered power law plus reflection, the model {\it pcref} described in
detail in Paper III. Positive Gaussians were added to represent the Fe K fluorescence line at $\sim$6.4 keV and an apparent red wing.

Initially restricting the fit to the 1.3-5 keV band, covering the K-shell resonance  lines of Mg, Si, S, Ar and Ca, the continuum model yielded
a moderately good fit ($\chi^{2}$ = 888/738). However, several absorption lines were clearly evident in the data:model ratio plot. The
addition of a single XSTAR component of grid 21 matched the stronger of these features and provided an excellent overall fit ($\chi^{2}$ =
743/735), with ionisation parameter log$\xi$=3.2$\pm$0.1 and column density 1.8$\pm$0.2$\times10^{22}$ cm$^{-2}$ (figure 3, top panel).  The
relative blue-shift of this more highly ionised component, 2.8$\pm$0.2$\times 10^{-2}$, corresponds to an outflow velocity of 9100$\pm$600 km
s$^{-1}$, consistent with the strongest  high velocity component found in NeX Lyman-$\alpha$ absorption in the RGS data (Paper I).  

Extending the EPIC spectral fit to 10 keV includes the strong Fe K absorption lines. The continuum was again modelled with {\it pcref}, but provided
a poor fit ($\chi^{2}$ = 1829/1358) with strong negative residuals near 7 keV. The addition of photoionised absorption with grid 21 produced a very 
substantial improvement ($\chi^{2}$ = 1445/1362), due mainly to matching the resonance absorption lines of Fe XXV and Fe XXVI (figure 3, mid
panel). The XSTAR fit over this wider energy band had a similar ionisation parameter (log$\xi$=3.4$\pm$0.1) and column density
(3.9$\pm$0.7$\times10^{22}$cm$^{-2}$), but a lower velocity (7600$\pm$500 km s$^{-1}$). A probable explanation for this difference, which appears to
go against the general velocity-ionisation trend, is that the Fe K absorption lines  are so strong as to dominate the fit, while the Fe
Lyman-$\alpha$ absorption line appears unusually broad in the data and may not be well modelled with a single velocity. The lower panel of figure 3 illustrates the relevant
structure of the Fe K absorption, where the ratio of data to continuum is modelled by Gaussians. It appears likely that the XSTAR fit is finding a mean
value in a velocity range from $\sim$5000 - 9000 km s$^{-1}$, perhaps tracing the post-shock flow across the main Compton cooling shell. In that
context, $\sim$5000 km s$^{-1}$ may then indicate the point where 2-body cooling becomes important as the temperature falls below $\sim$1 keV
(Section 6 and Appendix), and the highly ionised flow changes rapidly to a low ionisation flow. 

\begin{figure}                                                                                                
\centering                                                              
\includegraphics[width=6.2cm, angle=270]{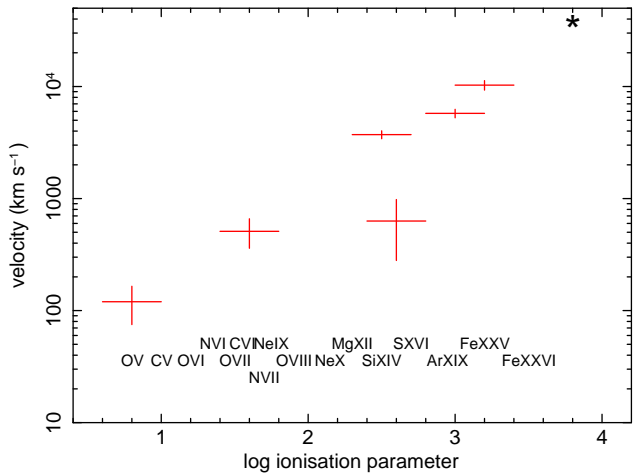}                                                                                  
\caption                                                                
{Outflow velocity and ionisation parameter for each of the XSTAR photoionised absorbers derived from fitting to the RGS 
spectra,  together with a high point representative of the pre-shock wind. The main trend is consistent with the linear correlation of velocity and
ionisation parameter expected for a mass-conserved  cooling flow}        
\end{figure} 

In summary, XSTAR modelling finds the RGS and EPIC absorption parameters are in good agreement in the high ionisation/high velocity region of
overlap.  The linear trend of velocity and ionisation parameter is clearly seen in figure 4, as expected for a cooling post-shock flow. The weak
high velocity component in the RGS data may indicate density structure in the post shock flow.

Finally, a 7th point is added to figure 4 to represents the pre-shock wind, with velocity v$\sim$0.12c (Paper III) and an ionisation parameter of
log$\xi$=3.8, the latter assuming the factor of 4 increase in density (along with a similar decrease in velocity) expected across a strong shock. 

An outstanding question for the mass-conserved shocked flow interpretation is raised by the strong column density gradient between the high and
lower ionisation XSTAR components. We interpret this in Section 6 in terms of the onset of additional two-body cooling, causing the flow to pass
quickly through intermediate velocity and ionisation stages. 

\section{Variable Fe K absorption and the highly ionised flow component}
The strongest absorption lines in the EPIC spectrum are generally those of the resonance transitions of FeXXV 1s-2p and FeXXVI Lyman-$\alpha$ (figure 3, mid panel). The ratio of
line depths is a sensitive measure of ionisation state, and a detailed examination of the Fe K absorption profile in Paper III showed that  the line
ratio - and implicitly the ionisation - could change significantly between adjacent orbits.

\begin{figure*}
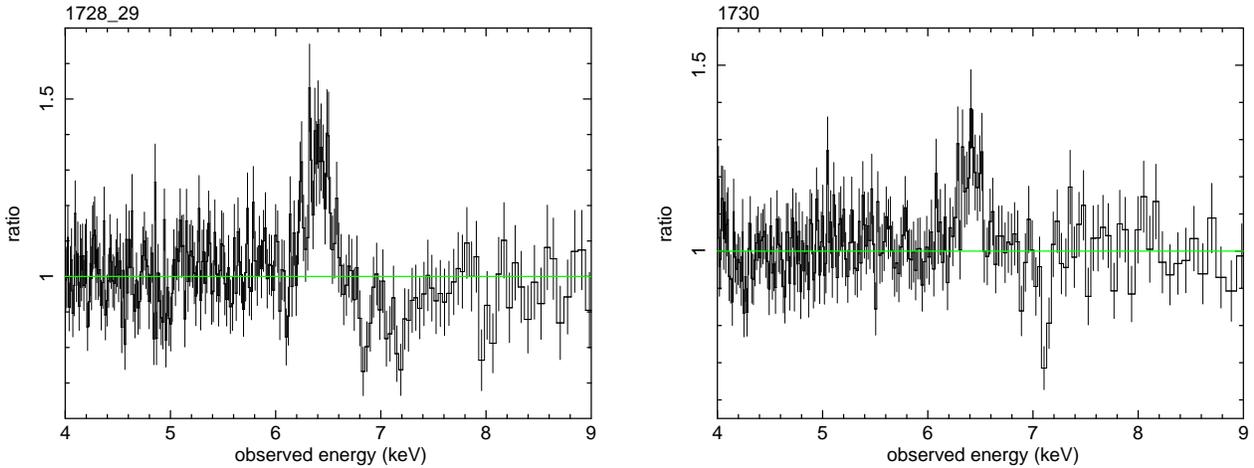

\begin{center}
\hbox{
 \hspace{0.5 cm}
   \includegraphics[width=6.2cm, angle=270]{1728_29.ps}
 \hspace{0.5 cm}
   \includegraphics[width=6.2cm, angle=270]{1730.ps}} 
   \end{center} 
\caption
{Variable absorption profiles for the composite of orbits 6 and 7 and for the peak flux orbit 8, indicating an increase in ionisation with higher
continuum flux} 
\end{figure*}

Figure 5 compares the Fe K absorption profile for orbits 6 and 7 (where similar spectra have been co-added for greater clarity)  with that for orbit
8 where the ionising flux $\ga$7.1 keV has increased by a factor of $\sim$2, to the highest level of the 2009  campaign. While the two absorption
lines are of similar depth in the first case, the 7.1 keV Fe XXVI Lyman-$\alpha$ line is significantly the stronger in orbit 8, indicating the
ionisation state of the highly ionised flow component has changed on a timescale of $\sim$2 days (table 1).

For an ionisation front moving into a low density medium, as here, the observed response time will be governed by the recombination time. Assuming a
post-shock temperature for the more highly ionised flow of $\sim$1 keV (Paper I), and a  recombination  coefficient of $\sim$$5\times10^{-12}$ 
cm$^{3}$ s$^{-1}$
(Verner and Ferland 1996), the observed variability timescale corresponds to an electron  density of  $\sim$$4\times10^{6}$ cm$^{-3}$. 

Comparing this measure of the particle density of the highly ionised flow with the corresponding column density from XSTAR modelling,
$N_{H}$$\sim$4$\times10^{22}$ cm$^{-2}$, yields a primary cooling shell thickness of $\sim$$10^{16}$ cm. We use this value in Section 6 to estimate
the shock radius.

\section{Recombination continua from the lower ionisation flow }

Paper II reported a significant emission component from the NVII RRC (threshold wavelength 18.59 \AA), though evidence for inter-orbit
variability is affected by blending with high velocity absorption in   OVIII Lyman-$\alpha$ (18.968 \AA).  The NVI RRC is strongly blended with OVII
absorption, while the OVII RRC sits at the upper wavelength edge of a strong Fe UTA. Fortunately, both CV and CVI RRC lie in spectral regions
relatively free from absorption lines.  Velocity profiles of both carbon RRC for a composite of all 15 orbits are of a similar form, with distinct -
though not identical variability - between individual orbits and groups of orbits with similar continuum levels.  The 15-orbit sum and a
representative set of emission profiles for CVI are shown in figure 6.

Figure 6 (top left panel) shows the composite velocity profile for the whole 2009 observation of \ngc, with zero velocity at the threshold
wavelength of the CVI RRC (25.303\AA). The statistical quality of the composite data, plotted at high velocity resolution (300 km s$^{-1}$), allows  non-RRC
features to be resolved and identified.  To the blue side they include the low velocity absorption component of NVI 1s-3p (seen near -5000 km
s$^{-1}$) and multiple NVII Lyman-$\alpha$ absorption components (near -7000, -10000  and -12500  km s$^{-1}$ in the plot).  Positive and
negative spikes at $\sim$8200 and $\sim$5000 km s $^{-1}$  are due to a dead pixel and a chip gap in the RGS CCDs.

\begin{figure*}
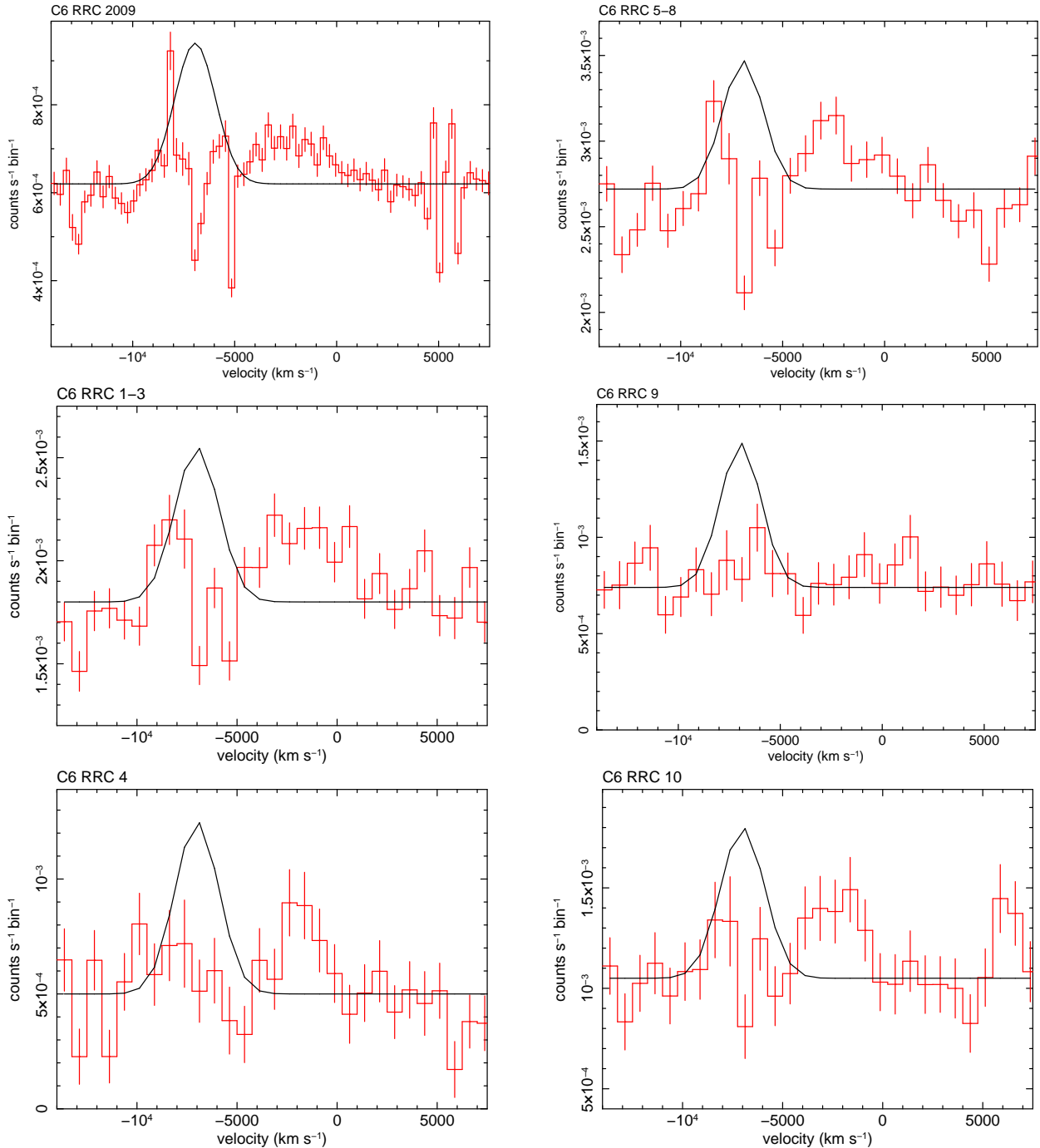
                                                                                                
\begin{center}
\hbox{
 \hspace{0.3cm}                                                             
 \includegraphics[width=6.1 cm, angle=270]{C6R_2009a.ps}
 \hspace{0.4cm}                                                             
 \includegraphics[width=6.1 cm, angle=270]{C6R_5-8a.ps}}
 \hbox{
 \hspace{0.3cm}                                                                  
 \includegraphics[width=6.1 cm, angle=270]{C6R_1-3.ps}                                                                                   
 \hspace{0.4cm}                                                        
\includegraphics[width=6.1 cm, angle=270]{C6R_9a.ps}} 
 \hbox{
 \hspace{0.3cm}                                                                  
 \includegraphics[width=6.1 cm, angle=270]{C6R_4.ps}                                                                                   
 \hspace{0.4cm}                                                        
\includegraphics[width=6.1 cm, angle=270]{C6R_10.ps}}
\end{center} 
\caption
{(left panel, from top) Composite velocity profile of the CVI RRC summed over the full 2009 observation. A strong RRC is visible, extending
blue-ward by $\sim$4000 km s$^{-1}$ from the rest frame threshold. The sharp spikes at -8200 and $\sim$5000 km s $^{-1}$ are due to a chip gap and
dead pixels in the RGS CCDs, while the absorption and emission structure beyond -4500 km s$^{-1}$ is identified with low
velocity absorption in the NVI K-$\beta$ line and the BEL and both low and higher velocity absorption in NVII Lyman-$\alpha$. The mid and lower panels show the same profile, with velocity resolution degraded from 300 to 750 km
s$^{-1}$, for orbits 1-3 and 4,  respectively. (right panel, from top) Composite velocity profile of the CVI RRC with velocity resolution of 750 km
s$^{-1}$, for orbits 5-8, 9, and 10. Note different y-axis scaling and see text for discussion.}        
\end{figure*}

The composite 15-orbit RRC  emission profile is quite unlike the saw-tooth shape expected for stationary matter, and extends blue-word of the zero
velocity  threshold by $\sim$4000 km s$^{-1}$. While some part of the RRC width will be a measure of the electron temperature in the recombining
gas, the shape of the composite profile indicates a mean outflow velocity of $\sim$2000 km s$^{-1}$, similar to the velocity gap at $\sim$1500-3500 km
s$^{-1}$ in the absorption profiles in the C, N and O Lyman-$\alpha$ lines (Paper I), and interpreted in Section 6 as a consequence of a rapidly
increased cooling rate. 

The middle and lower left panels of figure 6 show the same CVI RRC profile for a composite of the early orbits 1-3 and for orbit 4, when the mean
continuum level has fallen (figure 1) by a factor $\sim$4. The CVI RRC is of similar strength in both plots, setting a lower limit to the relevant
recombination time of $\sim$2 days (table 1). 

The right hand side of figure 6 continues the orbital sequence with the upper panel being a composite profile from the high flux orbits  5-8, with
the peak emission shifted still more strongly to the blue. In contrast, the high velocity RRC component has essentially disappeared  by orbit 9
(right side, mid panel), with an integrated  flux lower than for orbits 5-8 by a factor 4.5$\pm$1. 

Reference to the orbit timing data in table 1 shows an interval of $\sim$6 days between orbits 8 and 9, providing a firm upper limit to the
recombination time for the relevant flow component $\leq$5.5 days. However, additional  Swift and RXTE  monitoring (Alston \et\ 2013) shows a high
continuum  flux midway between orbits 8 and 9, further constraining the RRC decay time to $\leq$2.5 days. We assume below a recombination time of 2
days.

Assuming an electron temperature from the mean RRC profile of $\sim$4eV, a CVI recombination coefficient of $\sim$$10^{-11}$ cm$^{3}$ s$^{-1}$ (Verner
and Ferland 1996) and the  observed RRC decay time indicate a density of $\sim$$3\times10^{6}$ cm$^{-3}$.  The related column density  of
$\sim$$10^{21}$ cm$^{-2}$ from XSTAR modelling then translates to an absorbing path length of 3$\times10^{14}$ cm for the lower ionisation flow
component.  

Variability in the RRC contrasts with the lack of variability in the BEL of the same ions. The probable explanation lies in the higher 
blue shift of the RRC emission and a shorter light travel time.  Since a substantial RRC  flux implies a wide angle flow, a delayed response to a
lower continuum will be a product of light travel time and intrinsic recombination timescale, with the more blue-shifted emission from matter
observed at a smaller offset angle.  

A similar examination of the emission profiles of the CV RRC supports this conclusion, with the 15-orbit composite showing a clearer separation into
two velocity components. While the higher velocity component varies
in a similar way to CVI, the low velocity component in CV appears to vary little with overall flux level, similar to the BEL.

\section{Discussion}

Modelling the RGS and EPIC pn absorption spectra from the 2009 \xmm\ observation of \ngc\ confirms a general correlation of outflow velocity and
ionisation, consistent with mass conservation in a cooling shocked wind (King 2010, Paper I), while also quantifying the strong gradient in column
density with ionisation level. Inter-orbit changes in the ratio of FeXXV and FeXXVI absorption lines allow an estimate of the mean particle density in
the highly ionised  post-shock flow, with similarly rapid variability in strong RRC emission constraining the density in the lower ionisation flow. In
this Section we outline a model incorporating the above data, while providing a physical basis for identifying a fast ionised wind and a `warm absorber'
as early and late stages in a continuous, mass-conserved flow.

Figure 7 illustrates the envisaged  scenario, assuming spherical symmetry and a uniform radial flow.  The highly ionised wind collides with the ISM
sufficiently close to the AGN for strong Compton cooling of the shocked  gas to define a shell where the ionisation level remains sufficiently high for
ionised Fe K absorption. At a critical juncture along  the flow, 2-body processes become important (see Appendix) and the flow cools more rapidly over a
narrower region, where X-ray absorption (and emission) is dominated by the lighter metals (O,N,C).

For the highly ionised flow, a density estimate from observed variability in the Fe K absorption line ratio (Section 4), and an absorption column from
the XSTAR modelling indicate  an absorption length of $\sim$$10^{16}$ cm. Comparison with the theoretical Compton cooling  time for \ngc\ of  $\sim$600
R$^{2}$ yr (from Appendix, equ.2, with M=1.7$\times 10^{6}$$\msun$), and assuming a mean velocity for the highly ionised flow of 6000 km s$^{-1}$, finds
the observed cooling length to correspond to a shock at a  radial distance R $\sim$0.03 pc ($\sim$$10^{17}$ cm) from the black hole.

Repeated variability in the CVI RRC flux (Section 5) indicates a recombination time for the lower ionisation matter of $\sim$2 days. Together with a
related column density from XSTAR modelling, this indicates an effective absorbing path length of 4$\times10^{14}$ cm for the lower ionisation flow
component.  

Given these estimates of shell radius and thickness, the RRC emission measure provides an overall consistency check on the above scaling. A CVI RRC flux
from the spectral fit to orbits 5 - 8 of $\sim$$10^{-4}$ photons cm$^{-2}$s$^{-1}$ and recombination rate of $10^{-11}$ cm$^{3}$ s$^{-1}$, assuming 30 percent of
recombinations direct to the ground  state, corresponds to an emission measure of $\sim$$4\times10^{63}$cm$^{-3}$, for a Tully-Fisher distance to \ngc\
of 15.2 Mpc. 

For a mean density of $\sim$$3\times10^{6}$cm$^{-3}$, the emission volume (4$\pi.R^{2}$$\Delta$R) is then $\sim$$10^{50}$cm$^{3}$.  With the derived 
values of recombining shell thickness $\delta$R $\sim$$3\times10^{14}$cm, and shell radius R$\sim$$10^{17}$ cm, we find the measured RRC flux is
re-produced within a factor 2. Given the approximate nature and essential averaging  of many observed and  modelled parameters, the self-consistency 
check is remarkably good.

\begin{figure*}                                                                                                
\centering                                                              
\includegraphics[width=13cm]{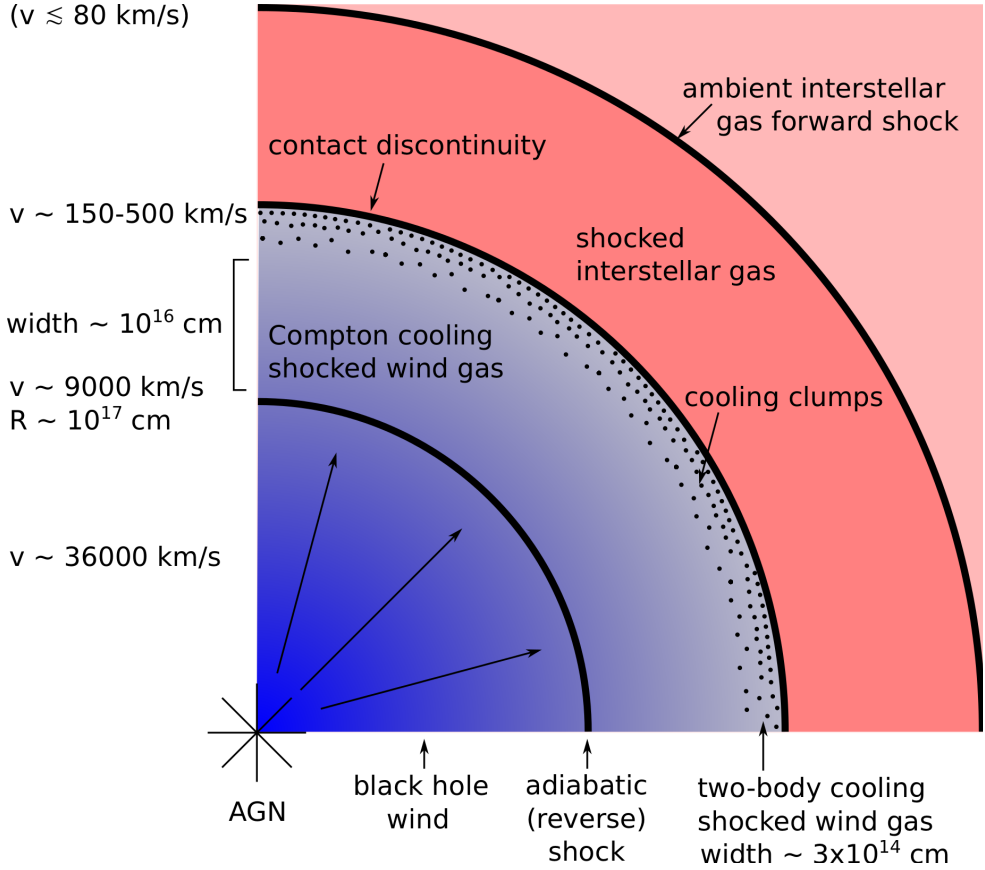}                                                                                  
\caption                                                                
{Structure of the \ngc\ outflow, not to scale, showing a highly ionised wind colliding with the ISM at a radius $\sim$$10^{17}$ cm, the strong shock
causing a sudden factor 4 drop in velocity. Strong Compton cooling of the shocked gas defines a thin shell where the velocity continues to fall but
the ionisation state remains sufficiently high for strong Fe K absorption. Further along, when 2-body processes become important, the flow cools
more rapidly and slows over a narrower region where absorption (and emission) are dominated by the lighter metals seen in the soft X-ray spectrum.
Although not used in the analysis, density variations in the shocked gas will be enhanced later in the flow and are pictured as cooling
clumps}        
\end{figure*} 

Finally, we note that all momentum-driven outflows in AGN will ultimately stall (King 2003, 2010), while the measured column densities in UFOs
show they must be variable on relatively short timescales (King 2010a). The repeated shocks will leave behind heated and compressed gas (both wind and ISM),
which can linger for some time before dispersing or falling back towards the black hole (note: the escape velocity for \ngc\ at R$\sim$$10^{17}$ cm 
is $\sim$500 km s$^{-1}$). 

We suggest that this matter, observed here as the low
velocity/low ionisation component, may form much of the warm absorber frequently observed in AGN (Tombesi \et\ 2013). The low ionisation, red-shifted
absorption component indicated in our XSTAR modelling offers an intriguing prospect of tracking the circulation of the stalled wind and the ISM that
has
been progressively removed from smaller radii.    

\section{Summary}

Combining the analysis of emission and absorption spectra with XSTAR modelling provides a picture of the overall \ngc\ outflow consistent with a
fast primary wind being shocked at a radial distance of order $10^{17}$ cm, well within the zone of influence of a supermassive black hole of mass
1.7$\times 10^{6}$ $\msun$. While the analysis necessarily involves substantial simplifications, with a few discrete components providing a coarse
description of a quasi-continuous and rapidly evolving flow, the resulting picture appears robust and physically attractive. 

The shocked gas initially cools in the strong radiation field of the AGN, with a Compton cooling length essentially determining the absorption columns
of  Fe and the other heavy metal ions. Two-body processes provide additional cooling as the density rises down-stream, rapidly becoming dominant.
Absorption (and emission) in the soft X-ray band then maps this thinner, outer region of the post-shock flow ahead of the contact discontinuity.

The inclusion of an unabsorbed continuum in the best-fit XSTAR model may be evidence for denser cooling clumps in the flow (figure 7). Although our analysis
has assumed a uniform radial flow, any density variations at the shock will be enhanced by differential cooling. Cooling radiation from the flow may also
contribute significantly to a partially covered soft X-ray continuum.        

We note that shocking with the ISM or other matter in the vicinity of the AGN, as found for \ngc, may provide the link beween ultra-fast
outflows (Tombesi \et\ 2010) and the equally common `warm absorbers' in such objects. In that scenario the onset of strong 2-body cooling
would  result in the intermediate column densities being relatively small and difficult to see with less sensitive observations than in the present
case. 

An important consequence of powerful AGN winds losing much of their energy by radiative cooling after shocking against the surrounding gas is that
feedback from such winds will be momentum-driven, consistent with the observed form of the observed M-$\sigma$ relation (King 2003).

\section*{Acknowledgements}  The work reported here is based on observations obtained with \xmm, an ESA science mission with instruments and
contributions directly funded by ESA Member States and the USA (NASA). Simon Vaughan was PI of the 2009 observation of \ngc. We thank the anonymous
referee for encouraging greater clarity in the text.

\section{Appendix}

Immediately after the (adiabatic) shock, slowing the fast wind from v$\sim$0.12c to 1/4 of this speed, 
the free--free and Compton cooling times are

\begin{equation}
t_{\rm ff} \cong 3\times 10^{11}{T^{1/2}\over N}~{\rm s}  = 2{R_{16}^2\over M_8\dot m}~{\rm yr}
\label{ff}
\end{equation}
and
\begin{equation}
t_{\rm C} = 10^{-4}{R_{16}^2\over M_8}~{\rm yr}
\label{compt}
\end{equation}
respectively (see King, Zubovas \& Power, 2011: here $T, N$ are the postshock temperature and number density, $R_{16}$ is the shock 
radius in units of $10^{16}$~cm, $M_8$ is the black hole mass in units of $10^8M_{\odot}$, and $\dot m \sim 1$ is the Eddington factor 
of the mass outflow rate). 

After the adiabatic shock, the gas cools rapidly from $T \sim 1.6\times10^{10}$~K by inverse Compton cooling, while its density rises 
as $N \propto T^{-1}$ (isothermal shock -- pressure almost constant). So 

\begin{equation}
t_{\rm ff}\propto {T^{1/2}\over N} \propto T^{3/2}, 
\end{equation}

which means that the free--free cooling time decreases sharply while the Compton time does not change. When $T$ has decreased enough 
below the original shock temperature, free--free (and all 
the other atomic two-body processes) becomes faster than Compton. 
 
From (\ref{ff}, \ref{compt}) above this requires

\begin{equation}
\left({T\over T_s}\right)^{3/2} < 5\times 10^{-5}
\end{equation}
or 
\begin{equation}
T < 2\times 10^7~{\rm K}
\end{equation}

showing the temperature of ionisation species forming around a few keV is determined by atomic cooling processes rather than Compton cooling. 
As the temperature falls further, two-body cooling becomes dominant, with the subsequent flow cooling still more rapidly and both velocity and ionisation 
level falling quickly. This strong cooling phase is traced in
the present observation by absorption and emission in the lighter metals (C, N, O) through increasingly narrow shells prior to the build up of low
velocity/low ionisation matter ahead of the contact discontinuity.

\end{document}